\documentclass[journal=jacsat,manuscript=article]{achemso}

\usepackage{chemformula} 
\usepackage[T1]{fontenc} 

\author{Jin Yao}
\affiliation{Institute of Electromagnetics and Acoustics, and Fujian Provincial Key Laboratory of Electromagnetic Wave Science and Detection Technology, Xiamen University, Xiamen 361005, China}
\author{Na Liu}
\affiliation{Institute of Electromagnetics and Acoustics, and Fujian Provincial Key Laboratory of Electromagnetic Wave Science and Detection Technology, Xiamen University, Xiamen 361005, China}
\author{Guoxiong Cai}
\affiliation{Institute of Electromagnetics and Acoustics, and Fujian Provincial Key Laboratory of Electromagnetic Wave Science and Detection Technology, Xiamen University, Xiamen 361005, China}
\email{gxcai8303@xmu.edu.cn}

\author{Qing Huo Liu}
\affiliation{Department of Electrical and Computer Engineering, Duke University, Durham, North Carolina 27708, USA}
\email{qhliu@duke.edu}

\title[An \textsf{achemso} demo]
  {Doubly Enhanced Third Harmonic Generation in Metal-Based Silicon Nanodisks}

\abbreviations{IR,NMR,UV}

\begin{document}

%
%
%
%
%

\begin{abstract}
Doubly enhanced third harmonic generation (THG) is realized by the electric dipole resonance (EDR) in silicon nanodisks placed onto a metallic film at near-infrared. By introducing the metal substrate, the perfect electric conductor (PEC) surface effect can be formed to not only enhance the electric field in silicon, but also improve the near-field distribution. Meanwhile, in consideration of the periodic nanostructure, the silicon nanodisk, acting as a high-refractive-index dielectric grating, can generate a propagating surface plasmon resonance (PSPR) on the metal surface. By flexibly manipulating the array period, PSPR can effectively couple with EDR, which produces further enhancement and novel properties for EDR. On account of the dual enhancements from these two effects, it is demonstrated that the total THG conversion efficiency of the proposed nanostructure is raised by more than eight orders of magnitude as compared with that of the all-dielectric nanostructure. In addition, the influence of silicon Kerr effect on the coupling and THG is investigated as well, manifesting an unprecedent THG efficiency around 10$^{-2}$ with input internsity 2 GW/cm$^2$. This work will pave a new way for improving the multipolar mode in metal-dielectric nanosturctures and facilitate its engineered applications in nonlinear optics, e.g. frequency conversion, spectroscopic and biochemical sensing.
\end{abstract}

\section{KEYWORDS}
harmonic generation; metal-dielectric nanostructures; perfect electric conductor surface effect; coupling effect; electric dipole resonance; propagating surface plasmon resonance.

~\\
Dielectric nanostructures have attracted tremendous attentions owing to their ability to efficiently confine and manipulate light within their volume, offering an attractive platform for various nonlinear processes,$^{1,2}$ e.g. harmonic generation,$^3$ wave mixing$^4$ and all-optical modulation.$^5$ Since the plasmonic devices suffer from Ohmic losses, heating detrimental and external local field, which restricts their nonlinear performance, dielectric nanostructures with low optical losses, high laser damage threshold and large internal mode volume have been considered as a candidate to overcome the current limitations of plasmonic nanostructures.$^{6-8}$

High-refractive-index low-loss semiconductors and dielectrics, such as AlGaAs, GaAs with second order susceptibility and Si, Ge with third order susceptibility, have emerged as alternative units for the design of nonlinear optics in all-dielectric nanostructures.$^{9-11}$ Based on the efficiently control of optically induced electric and magnetic Mie type resonances, Fano resonances and so on, the near field can be greatly concentrated inside the nonlinear dielectric materials and the corresponding nonlinear responses in such nanostructures can be substantially enhanced.$^{12-15}$ Recently, enhanced third-harmonic generation (THG) has been reported through a nonradiative anapole mode in germanium nanodisk,$^{16}$ which originates from the overlap of co-excited electric dipole (ED) and toroidal dipole (TD) moments.$^{17}$ Its conversion efficiency can be further boosted by introducing the plasmonic components to modify the local electric field, such as a surrounding metallic ring to form a plasmonic resonance around dielectrics,$^{18}$ and a metal substrate to generate a perfect electric conductor (PEC) surface effect inside silicon nanoresonators.$^{19}$ Therefore, as a key issue in metal-dielectric nanostructures, how to effectively combine the distrinct properities of both dielectric components and plasmonic components is demanded to be sloved by various attempts at structure designs, mechanism innovations and parameter optimizations.

Coupling effect is an interesting correlation between different modes when one mode approaches the other, which provides novel properties for both two modes. Many works have been reported for mode enhancement through coupling effects.$^{20-23}$ By adjusting the period to manipulate the coupling with propagating surface plasmon resonance (PSPR), a dramatic enhancement of the magnetic dipole resonance (MDR) was obtained.$^{24}$ The application of coupling effects in refractive index sensing by enhanced MDR was experimentally realized in ref 25, which shows a preeminent sensing performance close to the typical reports of electric resonances in plasmonic nanostructures. By enormously strengthening the local electric field and effectively optimizing the line shape of resonances, the coupling effect was also exploited to improve the optical bistability.$^{26}$ However, the coupling effects between the dielectric Mie type resonance and the PSPR in metal-dielectric nanostructures have not been explored, and the underlying mechanism of their nonlinear process still needs to be revealed.

In this work, we demonstrate an efficient THG with conversion efficiency $\sim$10$^{-2}$ from the proposed metal-based dielectric nanosturcture, as is shown in Figure 1. By exploiting the metallic substrate, two physical mechanisms are introduced to boost the near-field inside silicon. As a result, the maximal (relative) efficiency of doubly enhanced THG is more than two (three) orders of magnitude higher than that numerically and experimentally achieved from the same platform.$^{19}$ We firstly elucidate the process of double field enhancements originating from two effects, PEC surface effect and coupling effect, in the linear manner. Then, we simulate the nonlinear responses in both frequency domain and time domain, showing a preeminent THG efficiency at least eight orders of magnitude higher than that of SiO$_2$ substrate. At last, the influence of silicon Kerr effect on the proposed nanostructure is investigated as well. Further details regarding the dependence of pump pulse width on THG efficiency can be found in the Supporting Information.

\section{RESULTS AND DISCUSSIONS}



\begin{figure*}[tp]
\centering
{\includegraphics[width=3.5in]{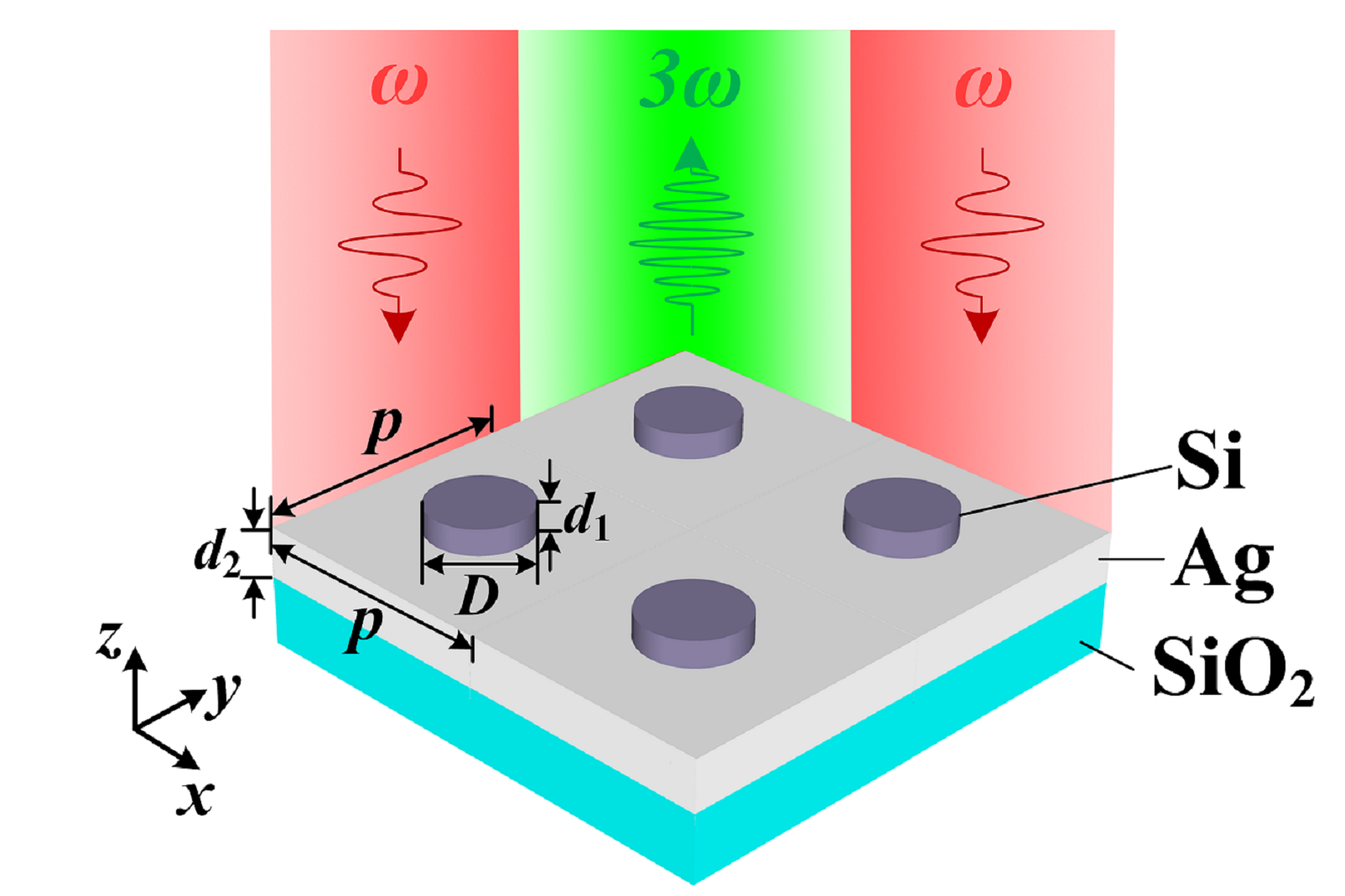}
\label{fig_first_case}}
\caption{Schematic illustration of the silicon nanodisk array on a silver film placed onto a SiO$_2$ substrate. Pumped by the fundamental plane wave with \emph{x}-polarization, third harmonic generation originating from the cubic optical nonlinearity of silicon can be efficiently excited. Geometric paramaters are diameter of silicon disks $\emph{D}$ = 400 nm, height of silicon disks $\emph{d}$$_1$ = 100 nm, thickness of silver film $\emph{d}$$_2$ = 200 nm, and \emph{p} is the array period.}
\end{figure*}

\textbf{PEC Surface Effect by Metal Substrate.} To gain the physical insight, we start with the scattering of a single silicon nanodisk placed on the silver film and analyze its PEC surface effect for field enhancement. With the \emph{x}-polarized plane wave normally illuminated from the air, Figure 2a and b give the multipolar decompositions (see Methods) of scattering cross-section (SCS) spectra for a silicon nanodisk placed on a silver film (silicon-on-metal (SOM) configuration) and an infinite SiO$_2$ substrate (silicon-on-dielectric (SOD) configuration), respectively. Diameter \emph{D} = 600 nm is adopted for the SOD configuration to ensure a similar EDR wavelength with SOM configuration. It can be seen that their total SCS spectra both exist a peak at wavelength $\lambda$ = 1410 nm, and they are mainly contributed by the ED moment, implying a EDR at this wavelength. Interestingly, benefiting from the metal, EDR in SOM configuration has a narrower line width, and its scattering efficiency is at least one order of magnitude higher than that in SOD configuration, leading to a higher resonance Q-factor and a stronger capability to manipulate the field.

\begin{figure*}[tp]
\centering
{\includegraphics[width=1.8in]{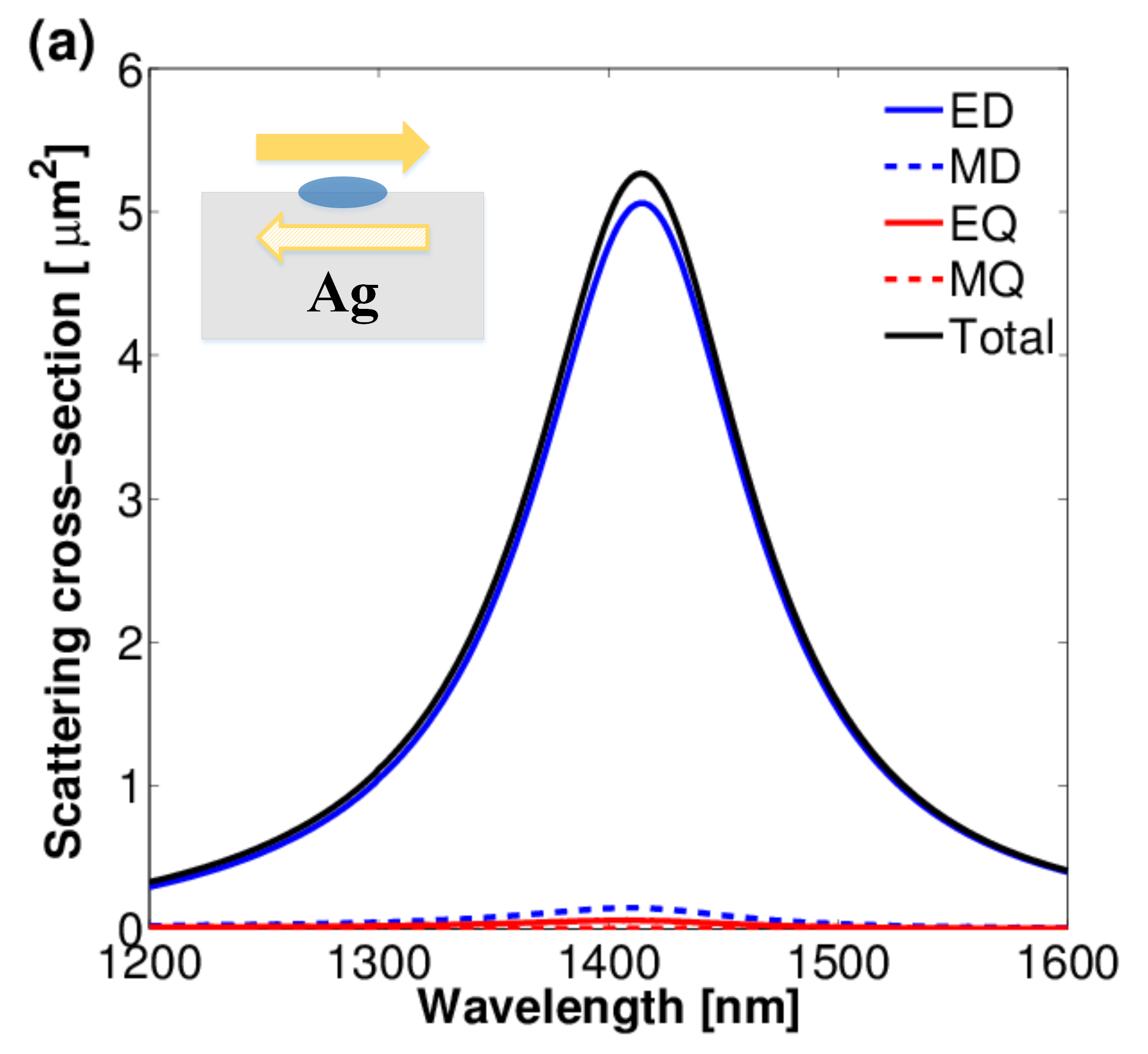}
\label{fig_first_case}}
\hfil
{\includegraphics[width=1.8in]{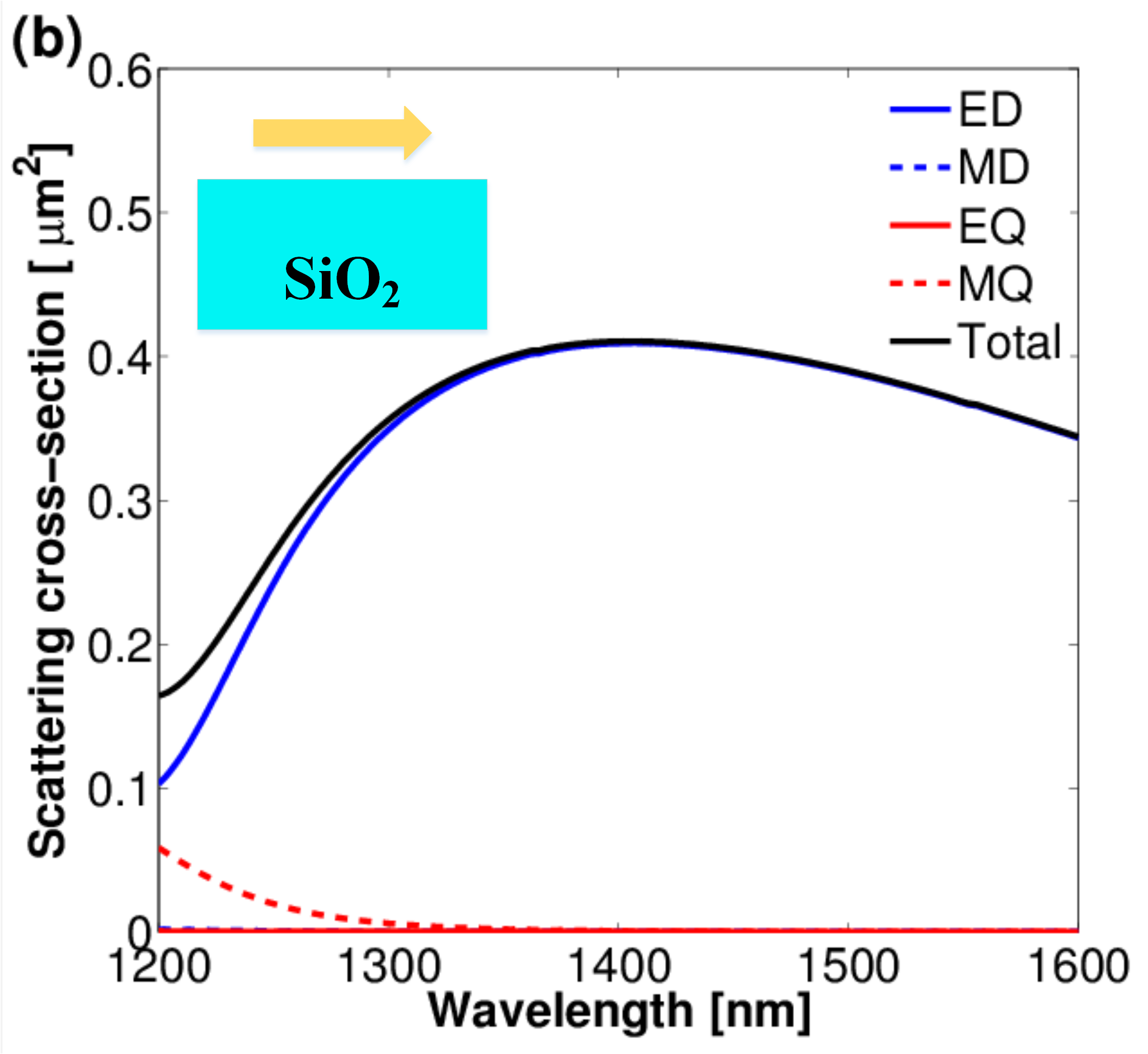}
\label{fig_first_case}}
\hfil
{\includegraphics[width=2.65in]{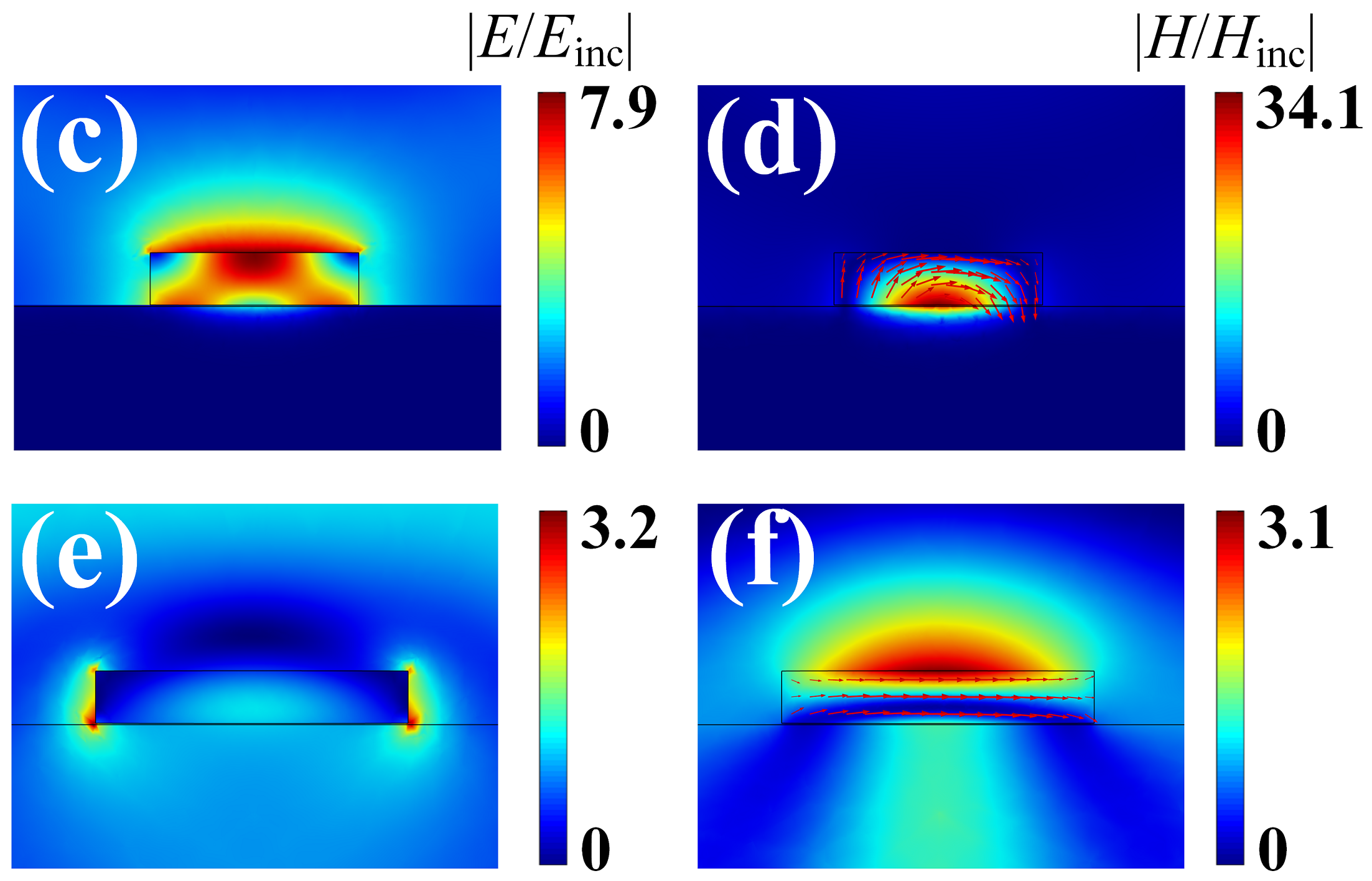}
\label{fig_first_case}}
\caption{(a) and (b) Multipolar decompositions of SCS spectra for SOM configuration and SOD configuration, respectively. The insets are schematic illustrations of currents. (c), (d) and (e), (f) Normalized \emph{xz}-plane electromagnetic field distributions at corresponding EDRs for SOM configuration and SOD configuration, respectively. The arrows denote the current orientations in the silicon nanodisk.}
\end{figure*}

Next, we elaborate the effect of a metallic mirror on electric currents. If a unidirectional current is placed on the PEC surface, the excited free electron oscillations will generate a virtual oppositely oriented image in metal.$^{19,27}$ An effective magnetic hotspot will thus emerge at the interface, further boosting the local magnetic field and thus the electric one. This process will induce a MDR for the entire configuration and a EDR for the silicon nanodisk,$^{28}$ which is depicted in the inset of Figure 2a. This PEC surface effect can be further confirmed by its current orientations and electromagnetic field distributions in Figure 2c and d. An intensive magnetic hotspot can be observed on the surface of metal, and the electric field and current orientations both show a half MDR (PEC-induced EDR) in the transparent region. Then we give the field distributions of SOD configuration in Figure 2e and f for comparison. It can be seen that two resonances are intrinsically different, the conventional EDR in SOD configuration can only focus its electromagnetic fields on the surface of silicon disk. However, for the SOM configuration with PEC effect, its local electromagnetic fields are both tailored and enhanced. A more concentrated near-field distribution inside the silicon nanodisk and a stronger field enhancement (|\emph{E}$_{\rm{max}}$/\emph{E}$_{\rm{inc}}$| = 7.8 and |\emph{H}$_{\rm{max}}$/\emph{H}$_{\rm{inc}}$| = 34.1) than those of SOD configuration can be attained in Figure 2c and d. Although the electric field enhancement inside silicon is not as strong as the magnetic one due to the nature of MDR, it still has the potential to provide an intensive nonlinear response.


\textbf{Coupling Effects between EDR and PSPR.} If the array period is taken into account, acting as a high-refractive-index dielectric grating on metal sustrate, the silicon nanodisk will induce a PSPR mode on the surface of silver. Figure 3a provides the absorption spectra of proposed periodic metal-based silicon nanostructures depending on the array period \emph{p}. Three valleys can be observed with changing the period. The bottom valley represents the [1,1]-order PSPR. Since it has little effect on other resonances at period \emph{p} = 1100 nm - 1800 nm, it will not be discussed for simplicity. 

The middle valley indicates the combination of [1,0]-order PSPR and the first-order Wood's anomaly, resulting in an asymmetrical lineshape instead of a classical Lorentz linetype, especially for \emph{p} < 1500 nm. According to their phase matching conditions$^{21,29}$
\begin{equation}
\frac{{2\pi }}{p}\sqrt {i_x^2 + i_y^2} + \frac{{2\pi }}{\lambda_0 }sin\theta  = \frac{{2\pi }}{\lambda_0 }\sqrt {\frac{{{\varepsilon _m}{\varepsilon _d}}}{{{\varepsilon _m} + {\varepsilon _d}}}} ,
\label{eq:refname2}
\end{equation}
\begin{equation}
\frac{{m\lambda_0 }}{p} = \rm{sin}\theta  \pm {\varepsilon _{\rm{d}}},
\label{eq:refname3}
\end{equation}
where $\lambda_0$ is the wavelength in vacuum, $\theta = 0$ is the incident angle, \emph{i$_x$} and \emph{i$_y$} are the diffraction orders in \emph{x} and \emph{y} directions, respectively, \emph{m} is an integer, and $\varepsilon_{\rm{m}}$ and $\varepsilon_{\rm{d}}$ are the relative permittivities of the metal and the dielectric in a surface plasmon polariton (SPP) waveguide, respectively. Due to the small volume of the silicon disk, $\varepsilon_{\rm{d}}$ is approximately equal to that of the air. According to eq 1 and eq 2, the resonant wavelengths of PSPR and Wood's anomaly can be both theoretically predicted with a direct proportional relation with the array period, and they will overlap rather than coupling with each other. Wood's anomaly is not further discussed either owing to its intrinsic weakness as compared with PSPR and little influence on the results. 

\begin{figure*}[tp]
\centering
{\includegraphics[width=3in]{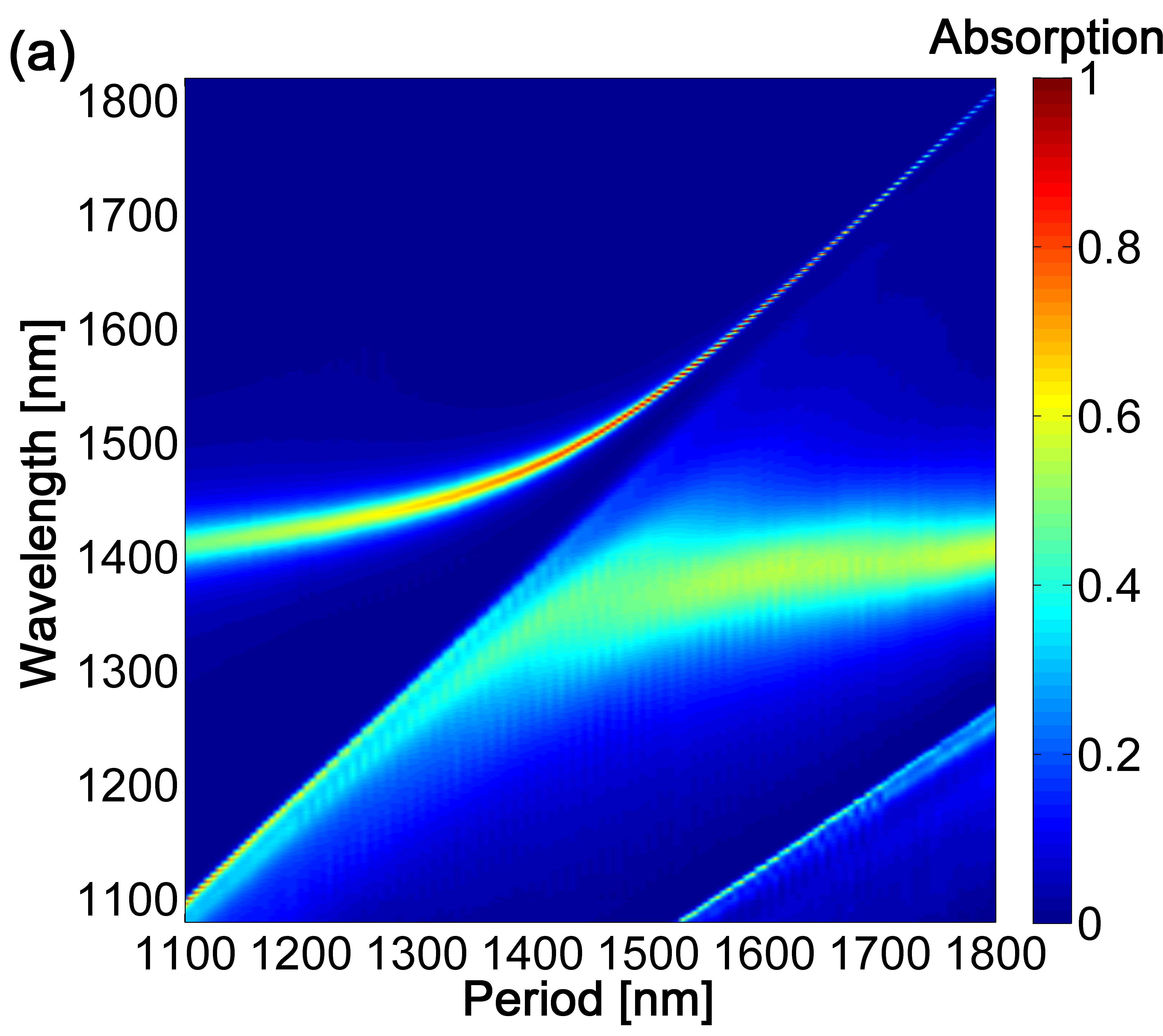}
\label{fig_first_case}}
\hfil
{\includegraphics[width=3in]{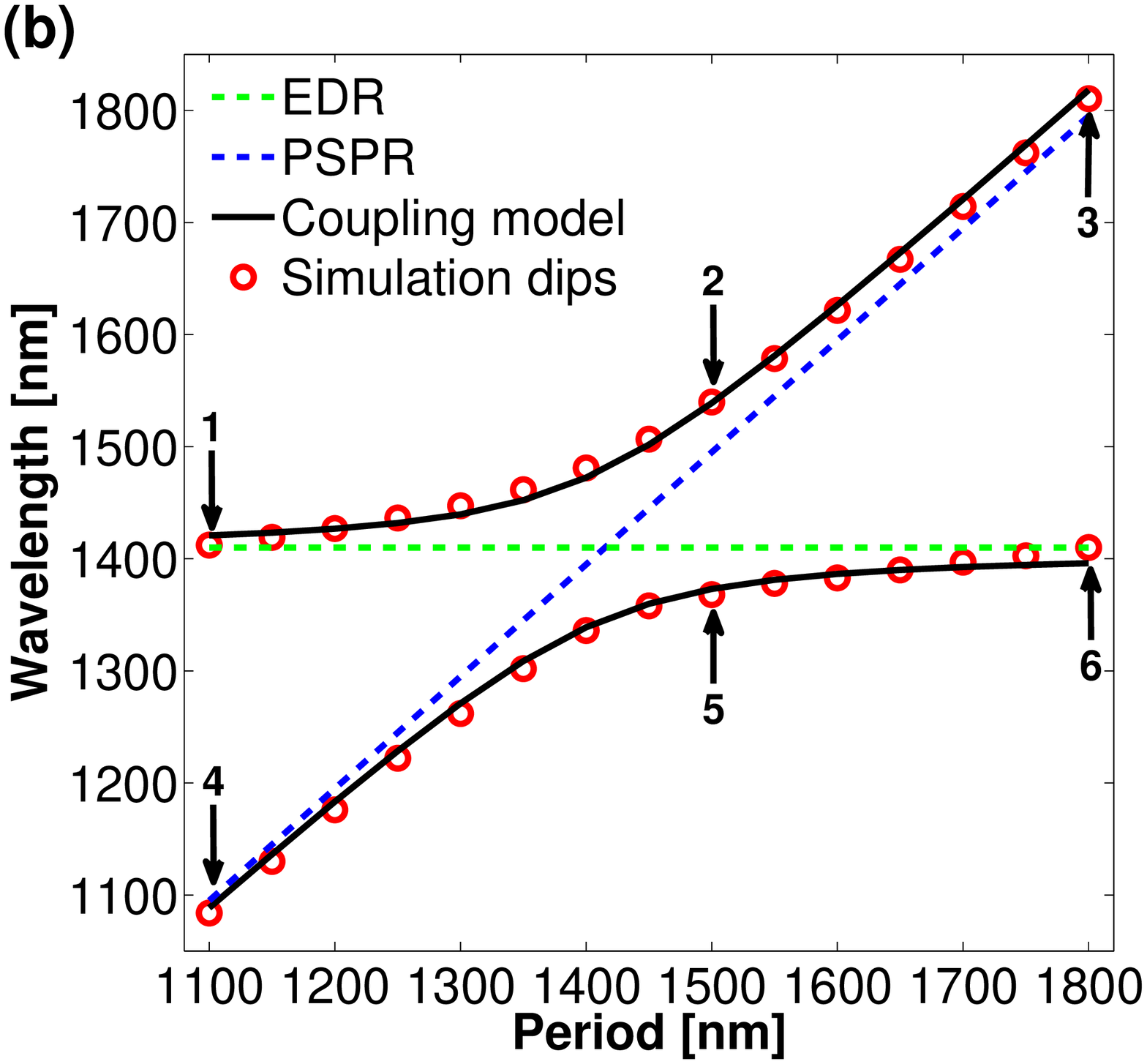}
\label{fig_first_case}}
\caption{(a) Absorption spectra of the proposed metal-based silicon nanostructures as the functions of the array period \emph{p}. (b) Dependence of resonant wavelengths on the period. Green and blue dashed line indicate the theoretical reosnant wavelengths of EDR and PSPR, respectively. Red circles are simulation results of two modes, which are extracted from the resonant wavelengths in Figure 3a. Black solid line is the fitting curve according to the coupling model.}
\end{figure*}

On the other hand, the top valley represents the EDR. Unlike PSPR, EDR mainly relies on the constituent material, the diameter and the height of the all-dielectric silicon nanodisks. If the period is larger than around 3$\times$diameter, the interaction between adjacent disks will be faint and little influence of period on EDR can be expected. Therefore, by manipulating the period, PSPR and EDR would approach and couple with each other. As can be seen in Figure 3a, with the period varying, the PSPR approaches the EDR, and these two resonances exhibit an interesting phenomenon of Rabi splitting.$^{30}$ In addition, due to PEC effect of the silver layer discussed above, the near-field profile of EDR is enormously improved, basically concentrated inside and around the silicon disks, which will produce a similar and overlapping field distribution with PSPR. It will further strengthen the coupling between two resonances and provide novel properties for both of them.

\begin{figure*}[tp]
\centering
{\includegraphics[width=3.65in]{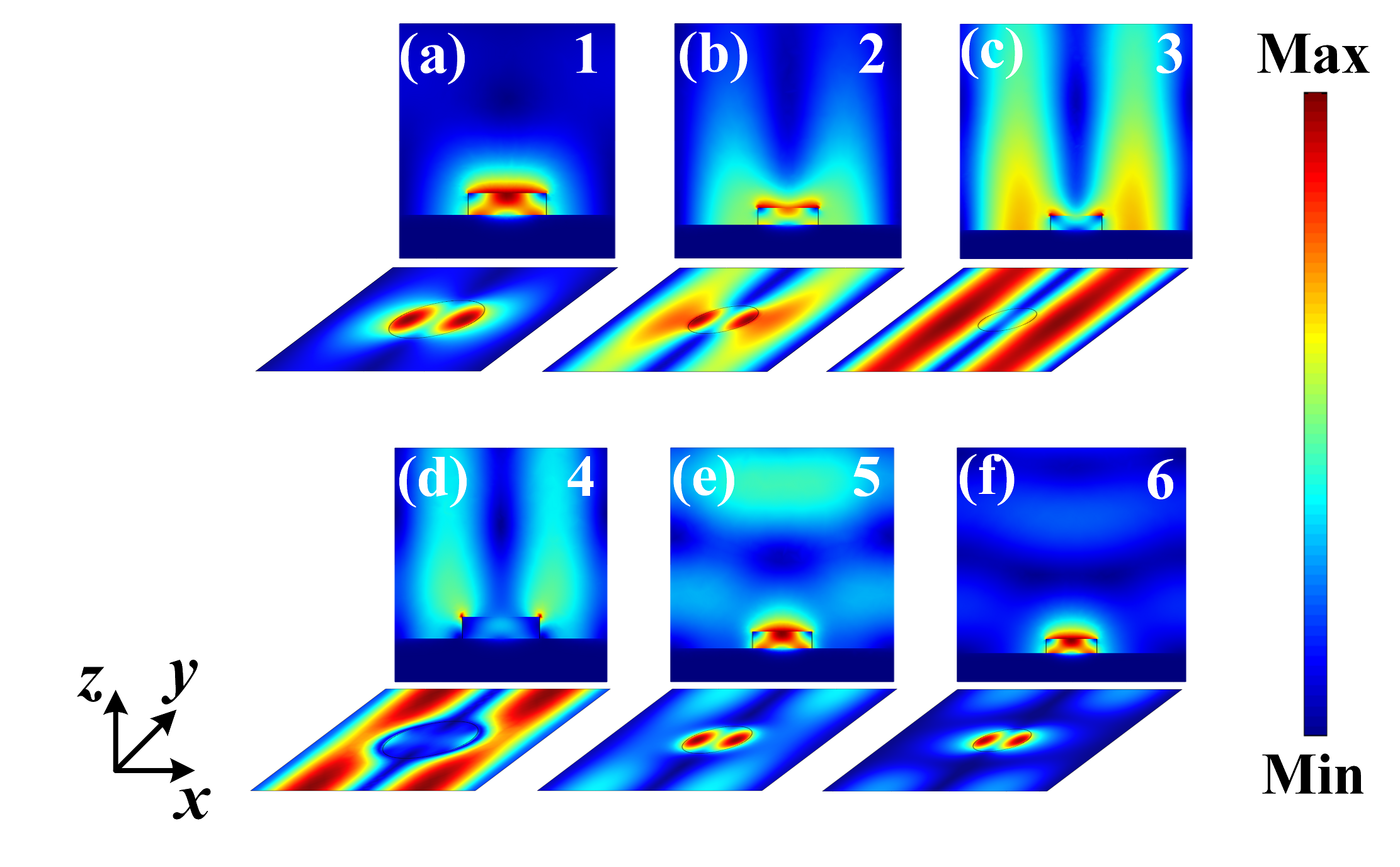}
\label{fig_first_case}}
\hfil
{\includegraphics[width=2.65in]{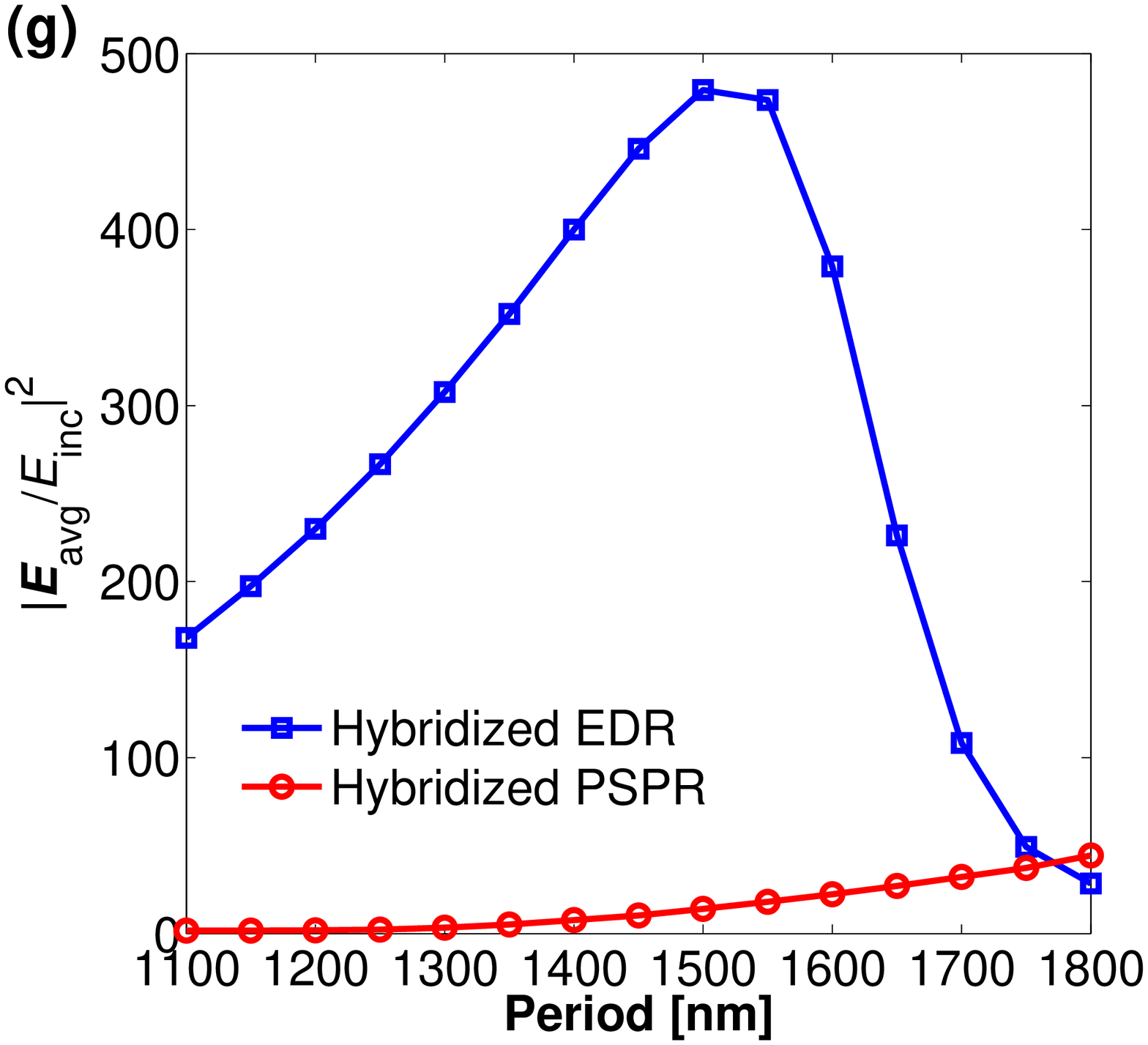}
\label{fig_first_case}}
\caption{(a) - (f) Normalized electric field |\emph{E}/\emph{E}$_{\rm{inc}}$| distributions of hybridized EDR and hybridized PSPR in both the \emph{xz}-plane and the \emph{xy}-plane with period \emph{p} = 1100, 1500 and 1800 nm, representing the status of un-coupling, strong coupling and over-coupling, respectively. Their corresponding resonant wavelengths are simulation dips 1-6 labeled in Figure 3b. (g) Dependences of average electric field enhancement |\emph{E}$_{\rm{avg}}$/\emph{E}$_{\rm{inc}}$|$^2$ of hybridized EDR and hybridized PSPR inside the silicon disk on the period. Results are extrated from the corresponding resonant wavelengths in Figure 3b.}
\end{figure*}

The anti-crossing of two resonant wavelengths is presented in Figure 3b. Dashed lines (green and blue) are the theoretical resonant wavelengths without coupling. Red circles indicate the resonance dips extracted from the simulation results in Figure 3a. It can be seen that two branches of simulation dips both deviate from the theoretical dashed lines due to the coupling effects between these two modes. Away from the strong coupling region around \emph{p} = 1500 nm, two simulation branches will follow approximately one of two theoretical lines. A coupling model is further utilized to describe the coupling as$^{30}$

 \begin{equation}
{E_{ + , - }} = \frac{{({E_{{\rm{EDR}}}} + {E_{{\rm{PSPR}}}})}}{2} \pm \sqrt {(\frac{{{{{E_{{\rm{EDR}}}} - {E_{{\rm{PSPR}}}}}}}}{2})^2 + (\frac{\Delta}{2})^2} ,
\label{eq:refname4}
\end{equation}
where \emph{E}$_{\rm{EDR}}$ and \emph{E}$_{\rm{PSPR}}$ are respectively the excitation energies of EDR and PSPR, and $\Delta$ stands for the coupling strength. By fitting the coupling model with the simulation results, we demonstrate a coupling strength is around $\Delta$ = 83.5 meV, which can basically predict two branches of resonant wavelength. This phenomenon of anti-crossing will broaden the way to effectively tune the resonant wavelengths by manipulating the period.

To vividly visualize the process of coupling effects, Figure 4a-f present the \emph{xz}-plane and the \emph{xy}-plane electric field distributions of hybridized EDR and hybridized PSPR extracted from the resonant points 1-6 labeled in Figure 3b. For the un-coupling region with \emph{p} = 1100 nm, the coupling between two modes is weak and they are provided with their intrinsic characteristics. As can be seen in Figure 4a and d, typical EDR and PSPR are effectively excited, respectively. The electric field of EDR is mainly concentrated inside and on the surface of the silicon grating, showing a dipolar distribution in the \emph{xy}-plane. PSPR focus its electric field at the interface between metal and dielectric, which is slightly perturbed by the silicon grating and rapidly attenuates in the \emph{z} direction. As the period \emph{p} increases to 1500 nm, due to the strong coupling in this region, two modes hold the similar electric field distribution that inherit both their features, as can be seen in Figure 4b and e. After the anti-crossing point, the process enters the over-coupling region. An external electric field similar to Figure 4d emerages in Figure 4c, and the field distribution in Figure 4f becomes more like that of EDR. Two resonances have exchanged properties with each other, indicating the gradual disappearance of the coupling. It should be noted that although the resonant properities have been significantly interchanged for Point 3, it is still designated as a hybridized EDR within the interseted period because it locates at the consecutive upper valley in the absorption spectra. So does Point 6.

According to Figure 4a-f, the electric field is significantly influenced by the coupling. Since it determines the performance of nonlinear response, we investigate the average electric field enhancement |\emph{E}$_{\rm{avg}}$/\emph{E}$_{\rm{inc}}$|$^2$  inside the silicon in Figure 4g.$^{26}$ It can be seen that hybridized PSPR is weak and its enhancement factor rises slowly with the increase of the period, so it is not suitable for the operation of nonlinear enhancement. For hybridized EDR, its enhancement factor attains its maximum of 479.3 at \emph{p} = 1500 nm, at least two times higher than that with un-coupling \emph{p} = 1100 nm and over-coupling \emph{p} = 1800 nm, which will lead to a preeminent nonlinearity at this period.

\begin{figure*}[tp]
\centering
{\includegraphics[width=3in]{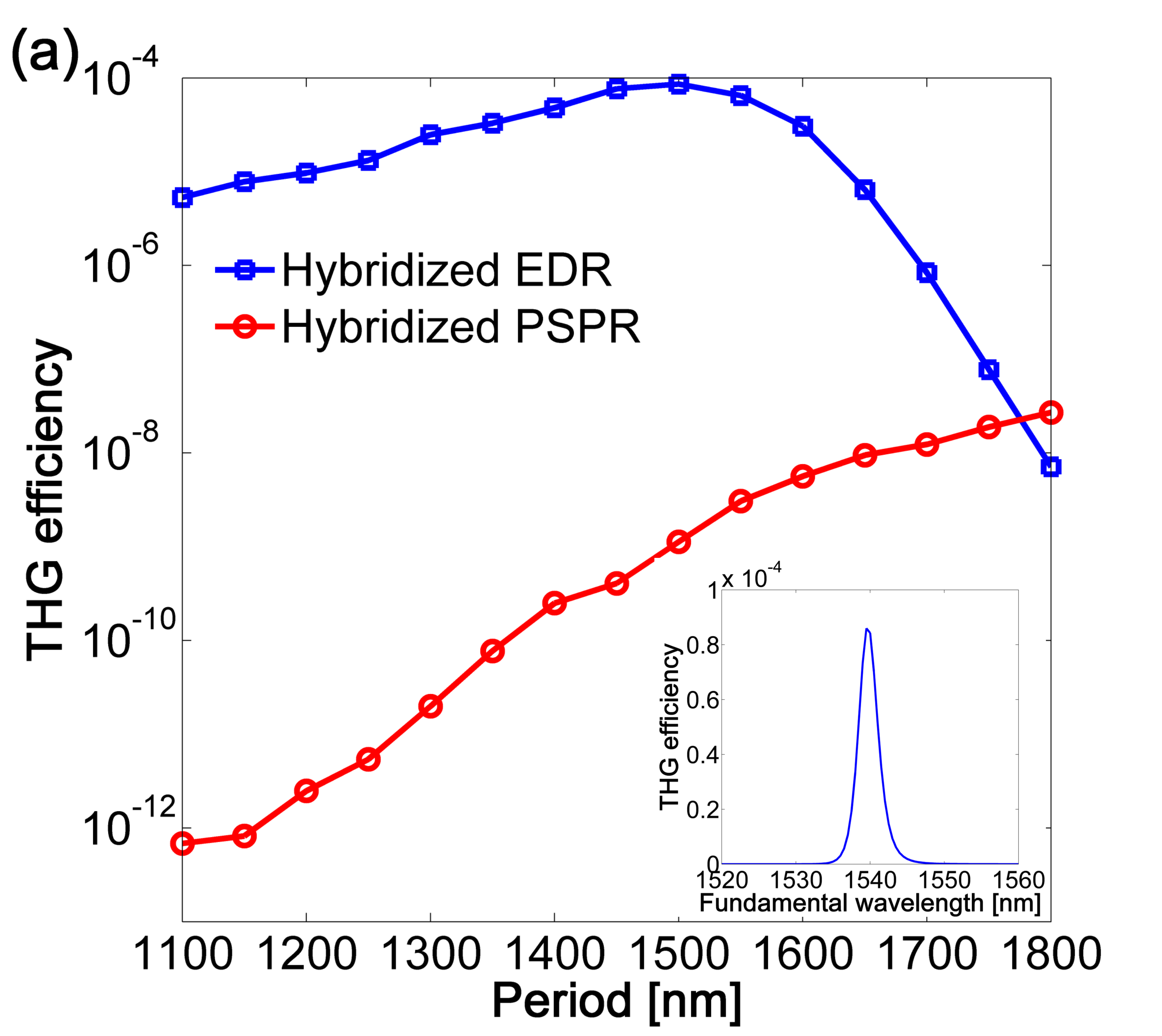}
\label{fig_first_case}}
\hfil
{\includegraphics[width=3in]{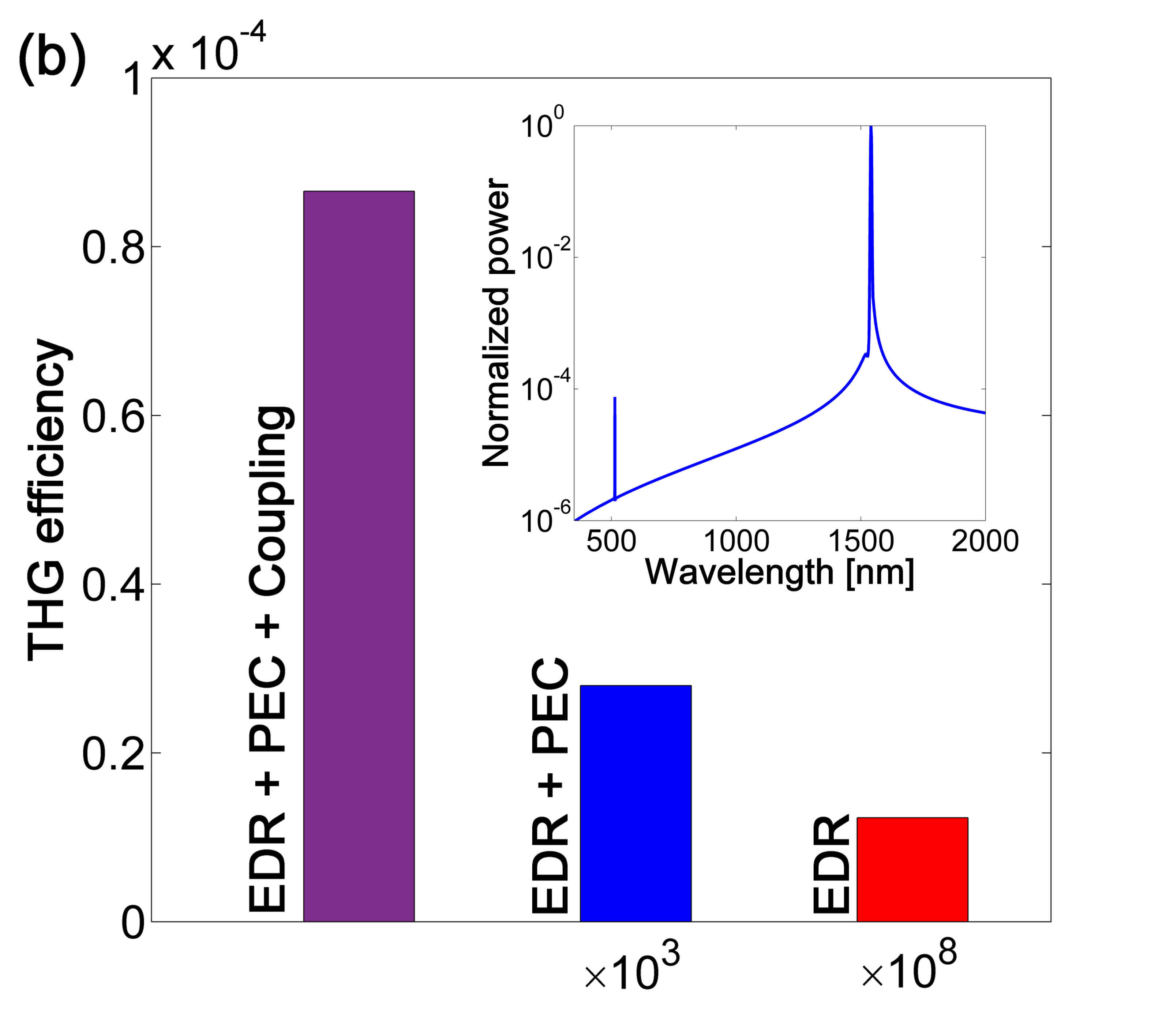}
\label{fig_first_case}}
\caption{(a) THG conversion efficiency as a function of the period. The inset is the nonlinear frequency domain spectrum of the hybridized EDR with \emph{p} = 1500 nm. (b) Nonlinear time domain responses of three cases, EDR with dual effects, EDR with only PEC effect and EDR with a SiO$_2$ substrate. The incident fundamental wavelengths are based on their corresponding resonant wavelengths. The inset shows the Fourier-transformed power spectrum of doubly enhanced EDR normalized to the incident pulse.}
\end{figure*}

\textbf{Enhanced Third Harmonic Generation.}  On account of the double enhancements from the above two effects, the nonlinear frequency domain reponses as functions of the period are thus investigated in Figure 5a. With a similar trend with the field enhancement, an optimal THG with efficiency 8.6 $\times$ 10$^{-5}$ (input intensity \emph{I}$_{inc}$ = 0.03 GW/cm$^2$) is acquired at \emph{p} = 1500 nm based on the hybridized EDR. Its nonlinear frequency domain spectrum is plotted in the inset of Figure 5a, showing a classical Lorentz line shape strongly relying on the resonance that is excited. If the incident wavelength detunes away from the resonant wavelength, the local electric field will be weakened, and the THG will gradually vanish. 

Next, the nonlinear time domain responses are studied in Figure 5b and its inset. It is demonstrated that two results in frequency domain and time domain can verify each other. THG efficiencies based on other two cases, EDR with only PEC enhancement and EDR without enhancement are given as well. In view of the comparison and the practical fabrication, EDRs with only PEC effect and with dielectric substrate are simulated as the periodic nanostructures with a sufficiently large period \emph{p} = 2500 nm instead of the single scattering nanoparticle. It can be seen that benefiting from the double enhancements, THG efficiency of 8.6 $\times$ 10$^{-5}$ is improved by more than three and eight orders of magnitude as compared with those of EDRs without coupling and with a SiO$_2$ substrate, respectively. Considering the low input intensity, the relative efficiency is three orders of magnitude higher than that obtained in the same platform,$^{19}$ and \emph{I}$_{inc}$ = 0.03 GW/cm$^2$ is the lowest pump intensity reported to achieve such a high efficiency.


\begin{figure*}[tp]
\centering
{\includegraphics[width=3in]{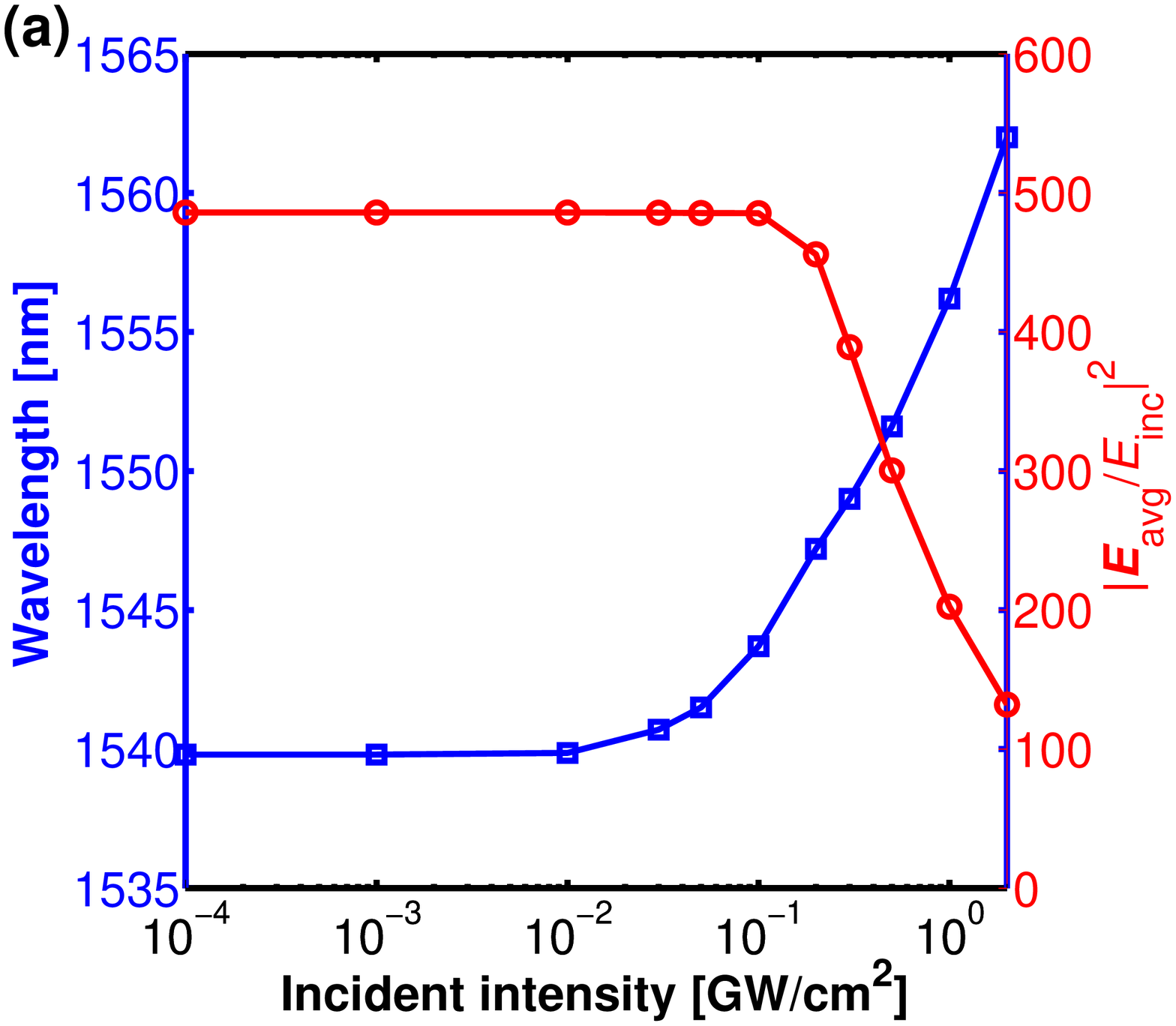}
\label{fig_first_case}}
\hfil
{\includegraphics[width=3in]{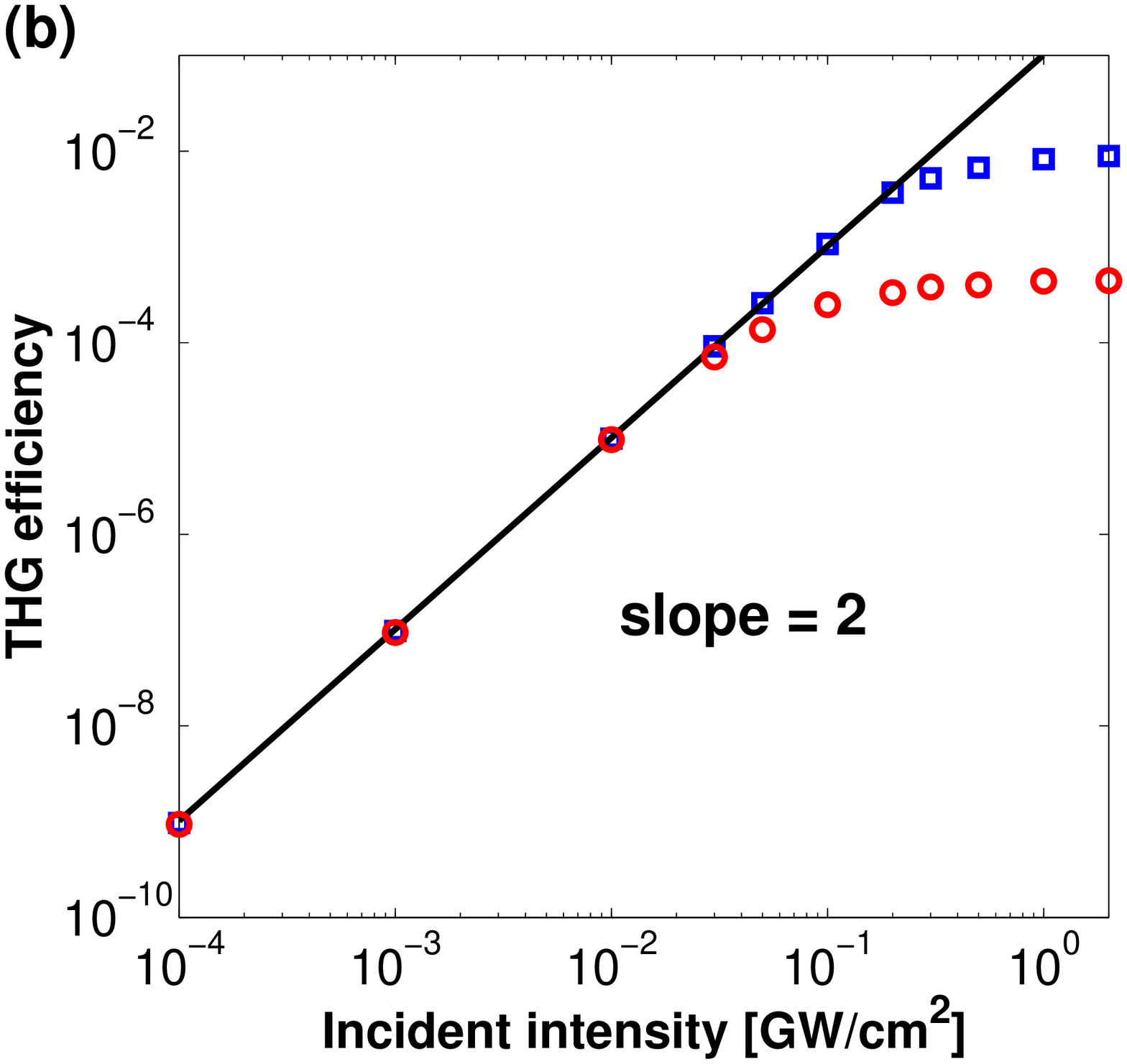}
\label{fig_first_case}}
\caption{(a) Dependences of resonant wavelengths and normalized average electric field enhancements |\emph{E}$_{\rm{avg}}$/\emph{E}$_{\rm{inc}}$|$^2$ on the incident intensity based on silicon Kerr effect. Results are extracted from the corresponding resonant wavelengths. (b) THG efficiencies as functions of the input intensity. Black straight line shows the theoretical predicted efficiency without silicon Kerr effect. Blue squares and red circles are simulation results extracted from the variable resonant wavelengths in Figure 6a and fixed wavelength $\lambda_0$ = 1540 nm, respectively.}
\end{figure*}

\textbf{Influence of Kerr Effect on Coupling and THG.} With the increase of incident intensity, Kerr effect, which perturbates the effective refractive index of the medium, cannot be neglected. Taking the silicon Kerr effect into account, Figure 6a provides the dependences of the resonant wavelength and the average electric field enhancement |\emph{E}$_{\rm{avg}}$/\emph{E}$_{\rm{inc}}$|$^2$ of the doubly enhanced EDR on the incident intensity. The resonant wavelength is increased with the boost of incidence and deviates more than 22 nm as incident intensity \emph{I}$_{inc}$ = 2 GW/cm$^2$, which is due to the high sensitivity of EDR to the refractive index of the silicon nanodisk. For PSPR, it is almost independent on the change of grating according to eq 1. Therefore, a disturbance to the coupling between two resonances will emerge. The attenuation of efficient coupling can be discovered according to the enhancement factor in Figure 6a. The enhancement remains constant before 0.1 GW/cm$^2$ and shows a rapid drop with \emph{I}$_{inc}$ > 0.1 GW/cm$^2$, indicating the coupling is rapidly weakened.

To investigate the influence of Kerr effect on nonlinear reponses, Figure 6b gives the THG efficiency depending on the incident intensity. Black straight line shows the theoretical predicted efficiency without silicon Kerr effect. Blue squares and red circles are respectively extracted from the corresponding resonant wavelengths in Figure 6a and fixed wavelength $\lambda_0$ = 1540 nm. The trend of THG efficiency with variable wavelengths is nearly identical to that of enhancement. Before 0.1 GW/cm$^2$, the THG efficiency remains a stable square relation with the input intensity. However, as the input intensity exceeds 0.1 GW/cm$^2$, THG efficiency manifests a slow rise and a growing departure from the proportional line, manifesting the increased damage of Kerr effect on the effective coupling. Its maximal THG efficiency is around 10$^{-2}$ with input intensity 2 GW/cm$^2$, at least two orders of magnitude higher than previous works about silicon nanostructures.$^{19,31}$ For the THG with fixed wavelength of 1540 nm, affected by both the weakened coupling and the changed resonant wavelength, its efficiency shows a faster deviation from the proportional line than that with variable wavelengths. It reaches the  maximum of 4.5$\times$10$^{-4}$ at \emph{I}$_{inc}$ = 2 GW/cm$^2$, which is still an unprecedent THG conversion efficiency, especially in such a simple periodic nanodisk structure.

\section{CONCLUSION}

In conclusion, we have utilized a metal film to generate the PEC surface effect and the coupling effect between two modes to doubly strengthen EDR in metal-based silicon nanostructures. By comparing different substrates, the enhanced electric field and the tailored near-field profile inside silicon disks can be obtained, resulting in an improvement of THG efficiency with at least five orders of magnitude. The conversion efficiency is further boosted by optimizing the coupling effects between PSPR and EDR, leading to an increase of more than three orders of magnitude as compared with that without coupling. Furthermore, the influence of silicon Kerr effect on the coupling and THG based on different incident intensities are also studied, attaining a maximal conversion efficiency around 10$^{-2}$ with input intensity 2 GW/cm$^2$. The results show that the THG conversion efficiency under double enhancements is at least two orders of magnitude higher than the previous works. This research will be interesting for the design of metal-dielectric nanostructures and promise prosperous applications in its enhanced nonlinearity.

\section{METHOD}
\textbf{Numerical Simulations.} Linear and nonlinear responses are both numerically simulated by commercial software COMSOL Multiphysics$^{\rm{TM}}$ based on the finite element method. Perfectly matched layers (PML) around the nanostructures are used to simulate the open space in scattering model and they are changed to periodic boundary conditions in \emph{x} and \emph{y} directions for periodic simulations. An incident $\emph{x}-$polarized Gaussian pulsed plane wave with the angular frequency $\omega  = \frac{{2\pi c}}{\lambda }$, where \emph{c} is the speed of light, defined by ${E_x} {\rm{ = }} {E_0}\cos (\omega t - {k_{\rm{0}}}z){e^{ - {{(\frac{{t - {t_0}}}{{\Delta t}})}^2}}}$, is exploited in time domain simulations. The peak amplitude of electric field is \emph{E}$_0$ = 1.504$\times$10$^7$ V/m, corresponding to an input intensity 0.03 GW/cm$^2$, and the pulse time delay \emph{t}$_0$ = 5 ps and the pulse width $\Delta$\emph{t} = 3 ps are chosen. The influence of pulse width on THG efficiency is discussed in the Supporting Information. For comparion and verification between frequency domain responses and time domain responses, the efficiency in time domain is amplified by 16$\sqrt{3}$ times to compensate the difference between Gaussian pulsed plane wave and plane wave and the influence of the narrowed THG pulse width. The refractive index of SiO$_2$ substrate is \emph{n}$_s$ = 1.5. The permittivity of the silver layer is described with experimental data.$^{32}$ The dispersive optical constant of Si is taken from the experimental data determined by ellipsometric measurements of hydrogenated amorphous Si.$^{5}$ It is chosen on account of a high nonlinear susceptibility at near-infrared.$^{33}$

THG is calculated based on an undepleted pump approximation, using two steps to obtain the nonlinear emission. We first simulate the field at the fundamental wavelength and obtain the induced nonlinear polarization in silicon. Then it is employed as a source term for computing the nonlinear response at the third-harmonic wavelength. The nonlinear efficiency is defined as the ratio of the total power of reflection and transmission divided by the incident power. The induced nonlinear polarization components in silicon at the THG wavelength can be calculated by$^{19}$
\begin{equation}
{{\textbf{P}^{(3)}} = {\varepsilon _0}{\chi ^{(3)}}(\textbf{E}{\cdot}\textbf{E})}{\textbf{E}} ,
\label{eq:refname4}
\end{equation}
where $\varepsilon _0$ is the vacuum permittivity, \textbf{E} is the localized electric field, and the third-order susceptibility tensor $\chi ^{(3)}$ is considered as a constant scalar value $\chi ^{(3)}$ = 2.45 $\times$ 10$^{-19}$ (m/V)$^2$.$^{34,35}$  The nonlinear polarization of silicon Kerr effect for the fundamental frequency field is denoted by$^{36}$
\begin{equation}
{\textbf{P}_{^{Kerr}}^{(3)} = 3{\varepsilon _0}{\chi ^{(3)}}{\left| \textbf{E} \right|^2} \textbf{E}} .
\label{eq:refname4}
\end{equation}

\textbf{Multipolar decomposition.} Multipolar decomposition of SCS is carried out based on light-induced polarization inside the nanoparticle ${\bf{P}}({\bf{r}}) = {\varepsilon _0}({\varepsilon _p} - {\varepsilon _s}){\bf{E}}({\bf{r}})$, where ${\varepsilon _p}$ and ${\varepsilon _s}$ denote the relative dielectric permittivities of the nanoparticle and the surrounding medium, respectively. ${\bf{E}}({\bf{r}})$ is the total electric field inside the disk, which can be obtained by the numerical simulations. The Cartesian multipole moments can thus be described as follow:$^{37,38}$
the Cartesian ED,
\begin{equation}
{{\bf{p}}_{{\rm{Car}}}} = \int {\bf{P}} {\rm{ d}}{\bf{r}}\\
\label{eq:refname4}
\end{equation}
the TD,
\begin{equation}
{{\bf{T}}_{{\rm{Car}}}} = \frac{{{k^2}}}{{10}}\int {\{ [{\bf{r}} \cdot {\bf{P}}]{\bf{r}} - 2{{\bf{r}}^2}{\bf{P}}\} } {\rm{ d}}{\bf{r}}
\label{eq:refname4}
\end{equation}
the magnetic dipole (MD),
\begin{equation}
{{\bf{m}}_{{\rm{Car}}}} =  - \frac{{i\omega }}{2}\int {[{\bf{r \times P}}]} {\rm{ d}}{\bf{r}}
\label{eq:refname4}
\end{equation}
the electric quadrupole (EQ),
\begin{equation}
{\widehat Q_{{\rm{Car}}}} = \int {\{ 3({\bf{r}} \otimes {\bf{P}} + {\bf{P}} \otimes {\bf{r}}) - 2[{\bf{r}} \cdot {\bf{P}}]\widehat U\} {\rm{ d}}{\bf{r}}} 
\label{eq:refname4}
\end{equation}
and the magnetic quadrupole (MQ),
\begin{equation}
{\widehat M_{{\rm{Car}}}} = \frac{\omega }{{3i}}\int {\{ [} {\bf{r \times P}}] \otimes {\bf{r}} + {\bf{r}} \otimes [{\bf{r \times P}}]\} {\rm{ d}}{\bf{r}}
\label{eq:refname4}
\end{equation}
where $\widehat U$ is the 3 $\times$ 3 unit tensor. Then, the total SCS $\sigma_{\text{scat}}$ of the disk can be calculated as

\begin{equation}
\begin{split}
{\sigma _{{\rm{scat}}}} \approx \frac{{{k^4}}}{{6\pi \varepsilon _0^2|{{\bf{E}}_{{\rm{inc}}}}{|^2}}}|{{\bf{p}}_{{\rm{Car}}}} + {{\bf{T}}_{{\rm{Car}}}}{|^2} + \frac{{{k^4}{\varepsilon _s}{\mu _0}}}{{6\pi {\varepsilon _0}|{{\bf{E}}_{{\rm{inc}}}}{|^2}}}|{{\bf{m}}_{{\rm{Car}}}}{|^2}\\
+ \frac{{{k^6}{\varepsilon _s}}}{{720\pi \varepsilon _0^2|{{\bf{E}}_{{\rm{inc}}}}{|^2}}}\sum {|{{\widehat Q}_{{\rm{Car}}}}{|^2}}  + \frac{{{k^6}{\varepsilon _s^2}{\mu _0}}}{{80\pi {\varepsilon _0}|{{\bf{E}}_{{\rm{inc}}}}{|^2}}}\sum {|{{\widehat M}_{{\rm{Car}}}}{|^2}}
\label{eq:refname4}
\end{split}
\end{equation}
where \emph{k} is the wavenumber and $\mu_{\text{0}}$ is the permeability of vacuum.

\begin{acknowledgement}

This work was supported in part by the National Natural Science Foundation of China (NSFC) (11604276, 61871340, 61871462), and National Key R$\&$D Program of China (2018YFC0603503).

\end{acknowledgement}

\begin{suppinfo}

Discussions on the choose of pulse width and its influence on THG efficiency (PDF).


\end{suppinfo}

\bibliography{achemso-demo}

\end{document}